\documentclass[manuscript]{aastex}
\pdfoutput=1 
\usepackage{natbib}
\usepackage{amsmath}
\usepackage{here} 
\usepackage{comment,color}
\newcommand{\rect}{\mathrm{rect}}
\newcommand{\supp}{\mathrm{Supp}}

\title{Fourth-order Coronagraph for High-Contrast Imaging of Exoplanets with Off-axis Segmented Telescopes}
\author{Satoshi Itoh\altaffilmark{1} and Taro Matsuo\altaffilmark{2,3}}
\altaffiltext{1}{Department of Earth and Space Science, Graduate School of Science, Osaka University,
1-1, Machikaneyama-cho, Toyonaka, Osaka 560-0043, Japan}
\altaffiltext{2}{Department of Physics, Graduate School of Science, Nagoya University, Furo-cho, Chikusa-ku, Nagoya, Aichi, 464-8602, Japan}
\altaffiltext{3}{NASA Ames Research Center, Moffett Field, CA 94035, USA}

\begin{abstract}
We propose a coronagraphic system with fourth-order null for off-axis segmented telescopes, which is sufficiently insensitive to the telescope pointing errors and finite angular diameter of the host star to enable high-contrast imaging of potentially habitable planets.
The inner working angle of the coronagraphic system is close to $1\lambda/D$, and there is no outer limit.
The proposed coronagraphic system is made up of a new focal plane mask and an optimized Lyot stop with the second-order null.
The new focal plane mask is  an extension of the band-limited masks with a phase modulation.
We construct a coronagraphic system with fourth-order null by placing two of the new coronagraph systems in succession to be orthogonal to each other.
 The proposed system is limited to narrow-band usage.
The characteristics of the proposed coronagraph system are derived analytically, which includes: (1)the leak of stellar lights due to finite stellar diameter and pointing jitter of a telescope, and (2)the peak throughput.
We achieve the performance simulations of this coronagraphic system based on these analytical expressions, considering a monochromatic light of 0.75$\mathrm{\mu}$m and off-axis primary mirror with a diameter of 8.5m. 
Thanks to the wide working area of the mask, the result shows that terrestrial planets orbiting K and G-dwarfs can be detected under the condition that the telescope pointing jitter is less than   $0.01\lambda/D\approx240$as. 
The proposed coronagraphic system is promising for detection of potentially habitable planets with future space off-axis hexagonally segmented telescopes.

\end{abstract}
\begin{document}
\maketitle
\newpage
\section{Introduction}
The data from the {\it Kepler space telescope} \citep{2017PAPhS.161...38B} has showed that about one-third of all M-dwarfs have  terrestrial planets \citep{2012ApJS..201...15H}. 
After the success of the {\it Kepler}, some terrestrial planets orbiting M-dwarfs in the habitable zone of their host stars have been found by ground-based telescopes \citep[e.g.,][]{2016Natur.536..437A,2016Natur.533..221G}.
The Transiting Exoplanet Survey Satellite  \citep[TESS; ][]{2015JATIS...1a4003R} is expected to detect such potentially habitable planet orbiting nearby M-dwarfs.
The next step is the characterization of the atmosphere of the potentially habitable planets using a spectroscope.  
 The prediction of features in planet spectra has attracted a lot of debate from an astrobiological perspective \citep{1986Natur.322..341A,2017ARA&A..55..433K,2018AsBio..18..709C,2018AsBio..18..619K,2018AsBio..18..630M,2018AsBio..18..663S,2018AsBio..18..739F}.
At mid-infrared wavelength, the non-thermal equilibrium of the atmospheres due to biological activity can be investigated from the thermal emission and transmission spectra.
The highly stable spectrophotometry \citep{2016ApJ...823..139M,2017AJ....154...97I,2018AJ....156..288G} and the nulling space interferometer \citep{1979Icar...38..136B,1997ApJ...475..373A,2011ApJ...729...50M} have been developed for this purpose.
In visible and near-infrared wavelength range, reflection spectra of potentially habitable exoplanets can be examined using high-contrast imaging with external occulter \citep{2000ApJ...532..581C,2006Natur.442...51C,2007ApJ...665..794V} or coronagraph.

Among these proposed methods, the most diverse is the coronagraphic methods \citep[e.g.,][]{1997PASP..109..815R,2000PASP..112.1479R,2002ApJ...570..900K,2005ApJ...622..744G,2005ApJ...633.1191M,2005OptL...30.3308F}. 
However, most of the coronagraphs operate for only monolithic pupils without discontinuities.
This non-ideality of the coronagraphic performance with respect to pupil discontinuities becomes a significant problem when these coronagraphs are applied to the segmented telescopes. 
This is because the planet-to-star contrast is $10^{-10}$ or $10^{-8}$ for G- or M-type dwarf in visible light.

Several researches have been carried out on the problem of pupil discontinuities to detect the potentially habitable planets. 
For example, \citet{2005ApJ...626L..65S} studied and examined the performance of a Lyot-type coronagraph for a monochromatic point source with off-axis apertures segmented by gaps of square lattice.
According to their study, the segmentation of telescopes limits the contrast to  $4g^2$ times the point spread function before coronagraph, where $g$ is the width of the gap normalized by the diameter of each segment.
\citet{2005ApJ...618L.161S} proposed the method of the apodized complex mask Lyot coronagraph (APLC).
This method theoretically provides the complete null of a point source in monochromatic light with the small inner working angle for any aperture shape.
 There have been several performance improvements in APLC designs for obscured, segmented apertures in the last few years \citep[e.g., ][]{2016ApJ...818..163N}.
\citet{2010ApJS..190..220G} proposed the application of the phase induced amplitude apodization \citep[PIAA; ][]{2005ApJ...622..744G} to the APLC.
This method is referred to as the PIAA complex mask coronagraph (PIAA-CMC). 
According to the result of the numerical simulation, an ideal PIAA-CMC appears to have second-order sensitivity to the telescope pointing jitter and the finite angular size of the host star \citep{2018SPIE10698E..1HB}.
\citet{2018JATIS...4a5004R} proposed a numerically optimized apodization pattern of a vortex mask coronagraph for off-axis segmented telescopes.
In their methods, an on-axis point source cannot be completely nulled across the focal plane.
In the numerical simulation, the stellar leakage can be swept out outside the angular radius of $ 20 \lambda / D $ to make a dark region on focal plane.
 The APLC, PIAA-CMC, and vector vortex coronagraphs each have manufactured prototypes and lab demonstrations in progress in broadband light \citep{2019BAAS...51g.101M}.
 In addition, the visible nulling coronagraph \citep[e.g., ][]{2016SPIE.9904E..20H} is another alternative coronagraph solution for segmented apertures.

In this paper, we propose a coronagraphic system with the fourth-order null. Our proposed system is effective for  off-axis segmented telescopes such as the Large UV/optical/infrared surveyor \citep[LUVOIR; ][]{2018SPIE10698E..0OB} and the Origins Space Telescope \citep[OST; ][]{2018SPIE10698E..15L}.
The proposed  coronagraphic system with the fourth-order null is sufficiently insensitive to telescope pointing errors and finite stellar diameter, enabling high-contrast imaging of potentially habitable planets.
Moreover, the inner working angle of the mask is close to $1\lambda/D$, and no outer edges exist.
In this paper, we define the inner working angles as off-axis angles at which the peak throughput reaches the first peaks from the inside.
The key element of the coronagraphic system is the new focal plane mask. 
The new  focal plane mask is  an extension of the Band-Limited Mask Coronagraph \citep[BLMC;][]{2002ApJ...570..900K}. The BLMC is a mask that controls only the modulus of the complex amplitude of light. This type of mask ideally works for only monolithic telescopes. However, the proposed coronagraphic mask introduces $\pi$ radian phase modulation to BLMC. Even when used with an unobscured segmented telescope, the stellar light on the re-imaged pupil after the new focal plane mask is completely nulled over the original pupil aperture.
 The proposed coronagraphic system is comprised of a new focal plane mask and a Lyot stop optimized for an unobscured segmented telescope. The coronagraphic system produces a second-order null.
The fourth-order version of this coronagraph is constructed by placing  the two second-order null coronagraphic systems in succession so as to be orthogonal to each other.
This concept of the proposed coronagraphic system with fourth-order null is ideally effective to detect the potentially habitable planets with future off-axis hexagonally segmented space telescopes.

In Section \ref{THE}, we present the concept of the proposed coronagraphic system which is optimized for the unobscured off-axis segmented telescopes. 
In Section \ref{SIM}, we use the analytical expressions derived in Section \ref{THE} to calculate the habitable planet signal and host star leakage on the new coronagraph mask assuming the LUVOIR-b architecture \citep{2019JATIS...5b5002G}. 
In Section \ref{DIS}, we discuss the extension of the new coronagraphic system to the polychromatic light and the prospect for manufacturing.
\section{\label{THE}Theory}
In this section, we review the principle of BLMC and describe a new coronagraph system for the unobscured off-axis segmented telescope.
The symbols used in this paper are summarized in  Table \ref{notation}.
\subsection{\label{CON}Concept}
\subsubsection{Band-limited Mask Coronagraph with Segmented Telescopes}
\begin{figure}[H]
\begin{center}
\includegraphics[width=170mm]{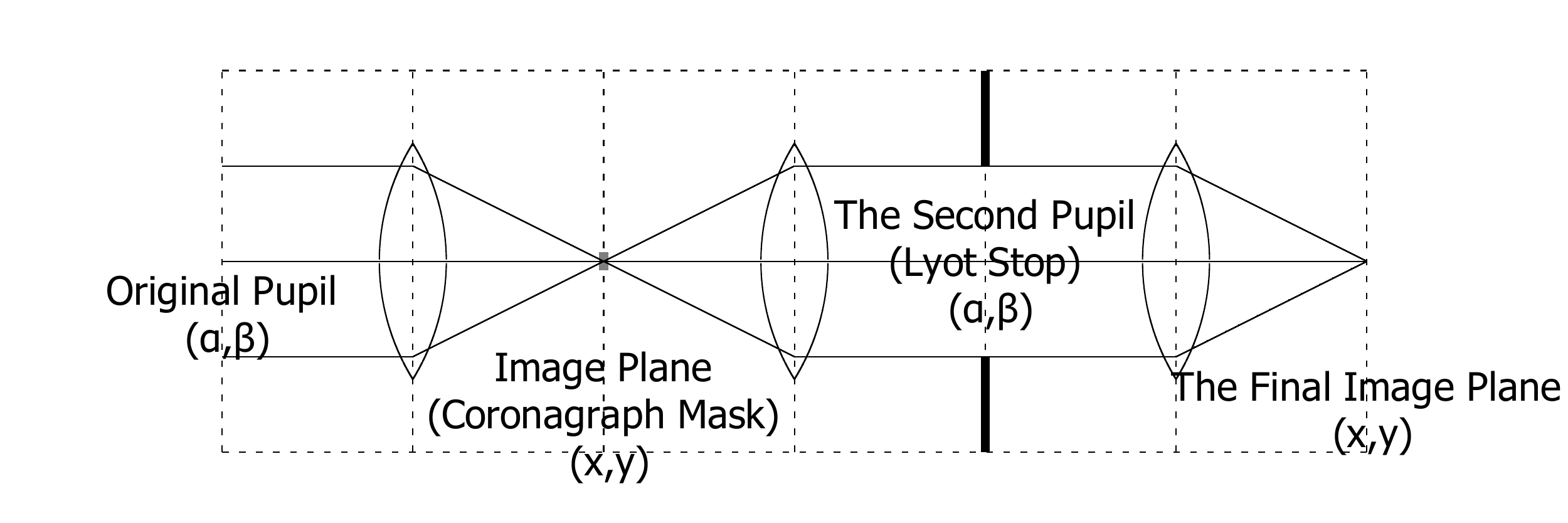}
\caption{\label{coordinate}Cartesian coordinates that are scaled by the imaging magnification of the pupil and image planes. The coordinates are normalized by $D$ and $\lambda/D$, respectively.}
\end{center}
\end{figure}

\begin{figure}[H]
\begin{center}
\includegraphics[width=160mm]{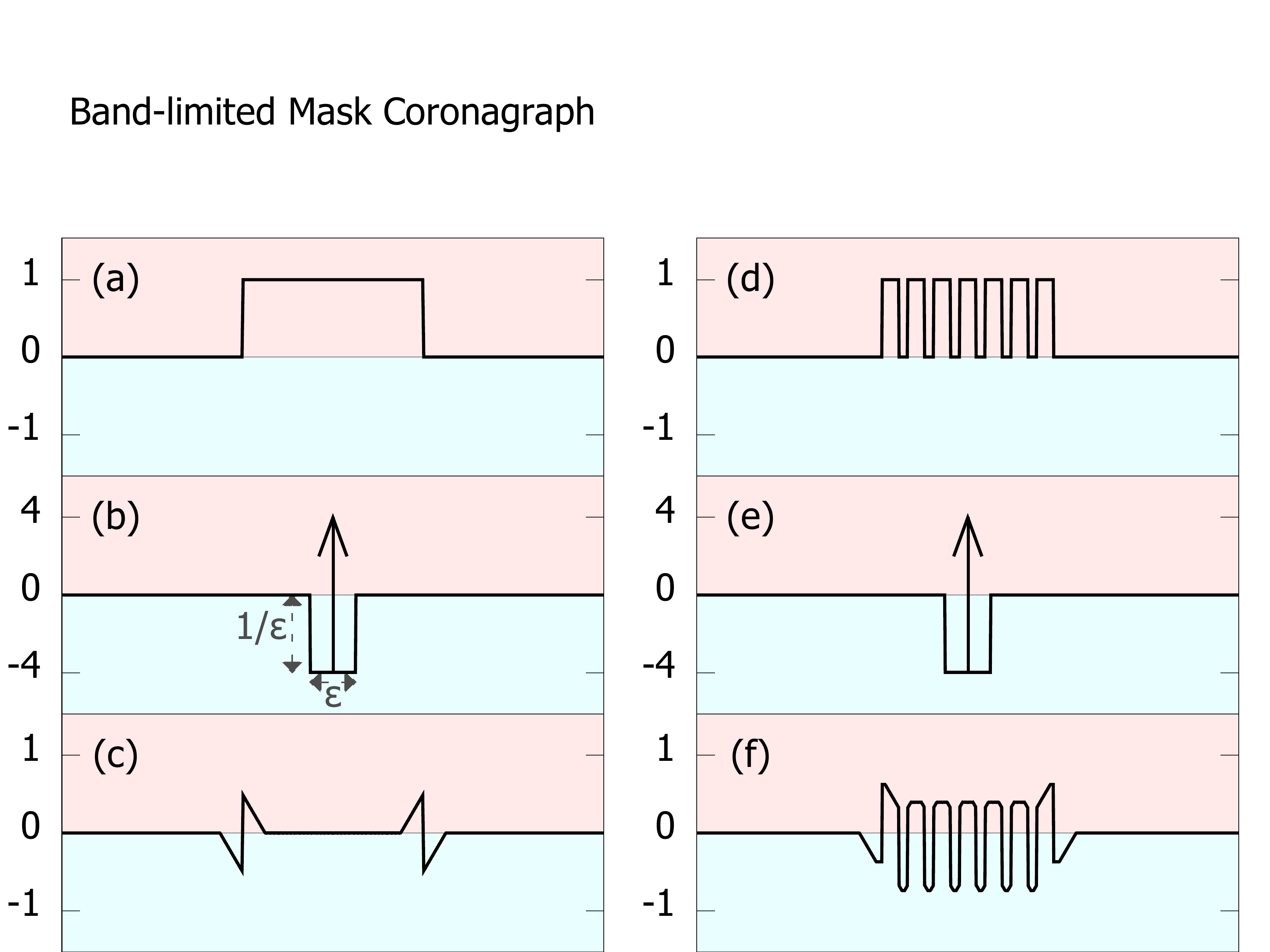}
\caption{\label{BLM}BLMC applied to monolithic (left column) and segmented (right column) telescopes. Each horizontal line represents zero amplitude and the vertical upward arrow means delta function. (a)A pupil function of a monolithic telescope. (b)Fourier conjugate of the BLMC mask function. (c)The second pupil amplitude (the convolution of (a) and (b)). (d)A discontinuous pupil function of a segmented telescope. (e)Fourier conjugate of the BLMC mask function. (f)The second pupil amplitude (the convolution of (d) and (e)). }
\end{center}
\end{figure}

Firstly, we review the BLMC for preparation. 
We define the Cartesian coordinates, $\vec{\alpha}=(\alpha,\beta)$ on the pupil and $\vec{x}=(x,y)$ on the focal plane, respectively, see Figure \ref{coordinate}. 
The coordinates are scaled to the imaging magnification and normalized by $D_{\alpha}$ or $D_{\beta}$ and $\lambda/D_{\alpha}$ or $\lambda/D_{\beta}$, respectively.
The mask function of the BLMC is generally written as:
\begin{equation}
M(\vec{x})=C(1-m(x)),
\label{M=C(1-m)}
\end{equation}
where $m(x)$ satisfies $m(0)=1\leftrightarrow \int_{-\infty}^{\infty}\! d\alpha\tilde{m}(\alpha)=1$. Also, $C$ is a constant and is determined such as the following two requirements are satisfied:
 (i)$|M(\vec{x})|\leq1$ (the assumption that amplitude masks cannot amplify light),
 (ii)$C$ is as large as possible for higher throughput of companion source (i.e., planet). 
The Fourier conjugate of $M(\vec{x})$ is as follows:
\begin{equation}
\tilde{M}(\vec{\alpha})=C\left(\delta(\alpha)\delta(\beta)-\delta(\beta)\tilde{m}(\alpha)\right).
\end{equation}
The second pupil amplitude is the convolution of $P(\vec{\alpha})$ and $\tilde{M}(\vec{\alpha})$:
\begin{equation}
P(\vec{\alpha}) * \tilde{M}(\vec{\alpha})=C\left( P(\vec{\alpha})-P(\vec{\alpha})* \delta(\beta)\tilde{m}(\alpha)\right ).
\label{convolution}
\end{equation}
Therefore, to null the second pupil amplitude at the point, $\vec{\alpha}$, we require the following condition:
\begin{equation}
\supp \left[\delta(\beta'-\beta)\tilde{m}(\alpha' -\alpha )\right] \in \supp\left[P(\vec{\alpha'})\right],  
\label{in}
\end{equation}
where $\supp\left[f(\vec{\alpha'})\right]$ means $\left\lbrace \vec{\alpha'}|f(\vec{\alpha'})\neq 0 \right\rbrace$. 
Considering Equation (\ref{in}), the BLMC can null the second pupil amplitude over a wide range of pupils with a monolithic telescope without apodization  (see (a)--(c) of Figure \ref{BLM}). However, the BLMC with a practical mask width over a segmented telescope cannot achieve a complete null on the second pupil due to the discontinuity of the segmented pupil (see (d)--(f) of Figure \ref{BLM}).
\subsubsection{Concept of   the New Coronagraph}
\begin{figure}[H]
\begin{center}
\includegraphics[width=160mm]{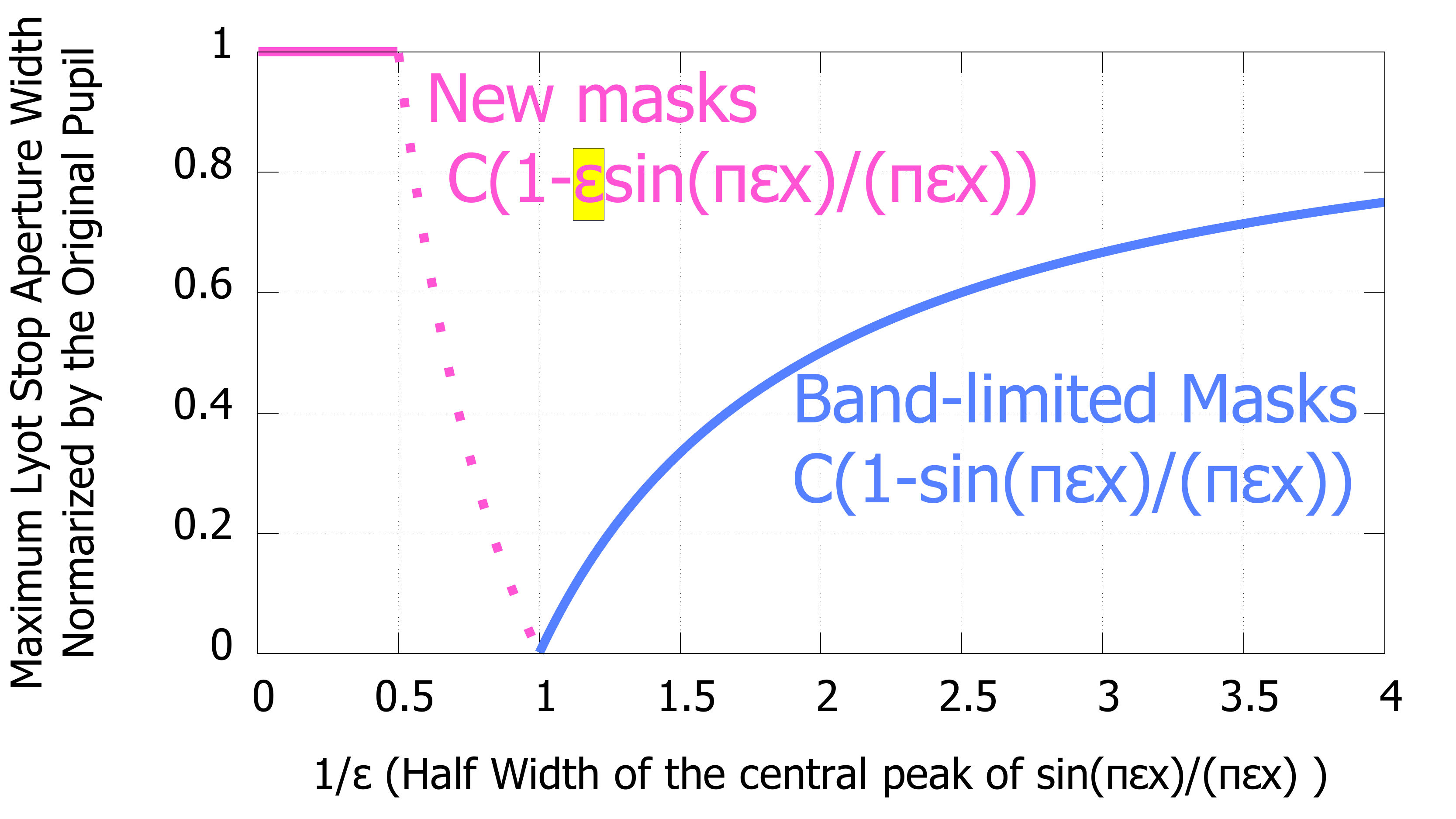}
\caption{\label{epsilon} Difference between the new mask and  BLMC (Maximum possible aperuture width of Lyot stop as a function of the width of the second term of the mask function). For simplicity, the discussion is limited to one dimension, and the original pupil function is assumed to be $rect[\alpha]$ in this figure. The width of the original pupil is 1. When $\epsilon^{-1}=1$, the width of the second term of the mask function and stellar diffraction amplitude are same. Note that the functional form at $0\leq\epsilon^{-1}<1$ (the new mask, the magenta line) and at $1\leq\epsilon^{-1}$ (the BLMC, the blue line) differ.  The maximum width of the Lyot stop aperture is same as the original pupil when $\epsilon^{-1}\leq0.5$ (the solid magenta line).  The ''$\epsilon$'' at the beginning of the second term of the new mask function works to add $\pi$-radian phase modulation onto the stellar diffraction amplitude. }
\end{center}
\end{figure}

To solve the above problem, we propose a new coronagraphic concept.
 Although the new mask type is an extension of the BLMC, the functional form and physical behavior of the new mask are qualitatively different from the BLMC (See Figure \ref{epsilon}).
 
\begin{figure}[H]
\begin{center}
\includegraphics[width=160mm]{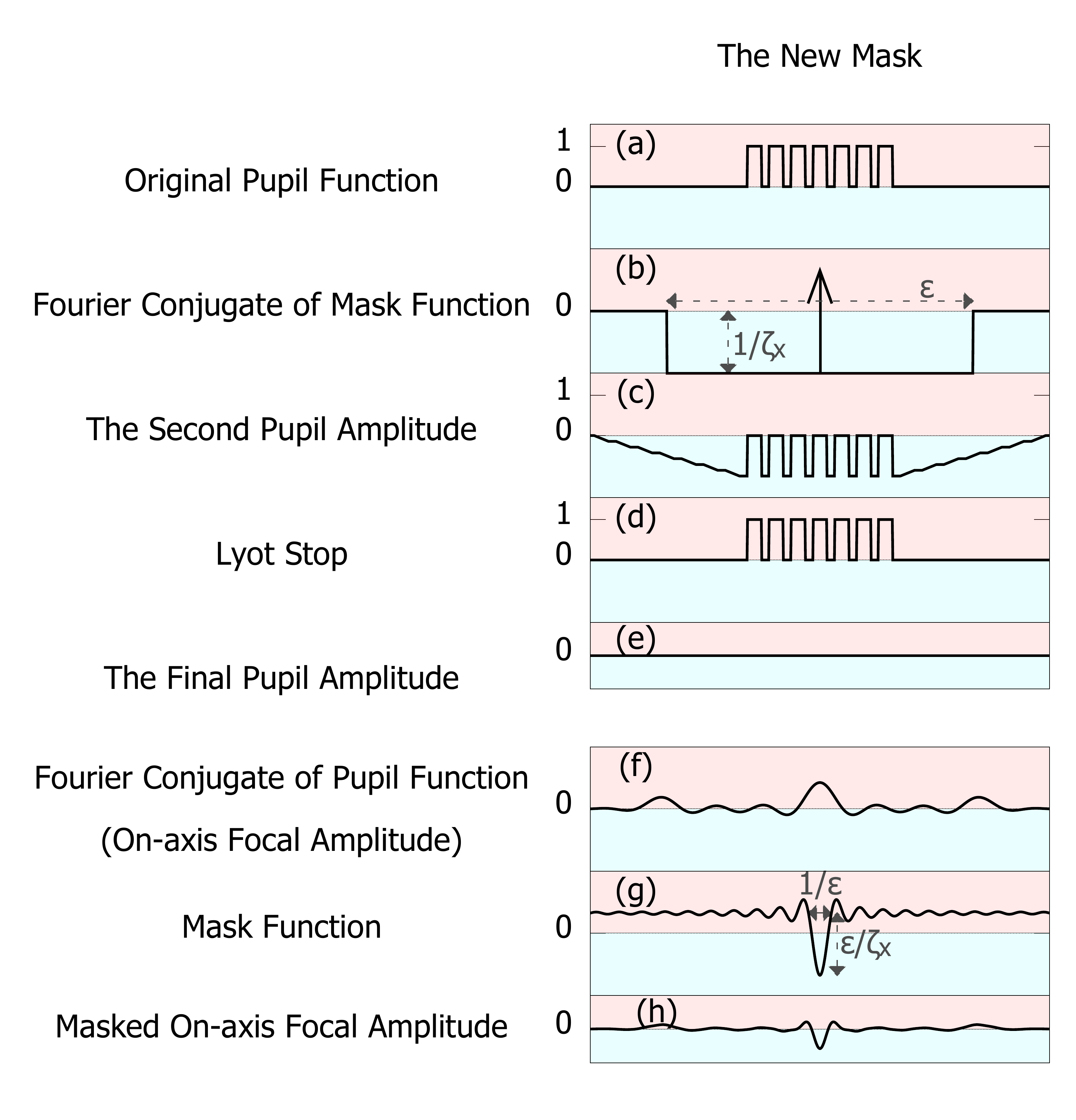}
\caption{\label{concept}Concept of the  new mask. Each horizontal line represents zero amplitude and the vertical upward arrow means delta function.  The constant factor, $C$, is omitted in this figure. (a)A discontinuous pupil function of a segmented telescope. (b)Fourier conjugate of the new coronagraphic mask function. (c)The second pupil amplitude (the convolution of (a) and (b)). (d)Lyot stop aperture function. (e)The final pupil amplitude. (f)Fourier conjugate of pupil function (on-axis focal amplitude before masking). (g)The new function. (h)Amplitude of on-axis source modulated by the new mask on focal plane (the product of (f) and (g)).}
\end{center}
\end{figure}

One of the BLMC mask pattern,  $m(\vec{x})=\frac{1}{\epsilon}\frac{\sin( \epsilon\pi x)}{\pi x}$, is the starting point for  derivation of this concept, where $\epsilon$ is a real number.
We choose this pattern because the $\tilde{m}(\vec{\alpha})$ is a rectangular function,  $\frac{1}{\epsilon}\rect[\epsilon\alpha]$, where $\rect[\alpha]$ takes a constant value of $1$ at $|\alpha|<0.5$, $0.5$ at $|\alpha|=0.5$, and $0$ at $0.5<|\alpha|$. 
The other BLMC mask patterns are not  to be used to derive  the new coronagraph concept.
The important point is that $P(\vec{\alpha})*\tilde{m}(\vec{\alpha})$  is a constant over the original pupil aperture when the following two conditions are satisfied: (1)$\supp\left[\tilde{m}\right]$ is two  or more times  wider than the original pupil aperture (see Figure \ref{concept}, (b)), (2)$\int_{-\infty}^{\infty}\!\! d\alpha P(\vec{\alpha})/P(\vec{\alpha})$ is a constant (hereinafter referred to as $\zeta_x$) at the original pupil aperture;  for exampe, when the pupil is simlply $\rect[\alpha]$ which has no gaps, $\zeta_x$ takes $1$. 
When simply setting $\epsilon$ of the mask function of the BLMC to $2$ or larger, the value of $P(\vec{\alpha})*\tilde{m}(\vec{\alpha})$ on the original pupil aperture becomes a constant, $\zeta_x\epsilon^{-1}$.
Since this constant is not $1$, the pupil amplitude on the second pupil cannot be nulled.

However, by multiplying $m(\vec{x})$ by a constant factor such that the value of  $P(\vec{\alpha})*\tilde{m}(\vec{\alpha})$ is $1$, the mask can null the pupil amplitude at the original pupil of the second pupil.
Therefore, the solution to derive the new mask is redefinition of $m(\vec{x})$ by multiplying $\zeta_x^{-1}\epsilon$ for $m(\vec{x})$  above:
\begin{eqnarray}
m_{\mathrm{new}}(\vec{x})&=&\zeta_x^{-1}\epsilon \times m(\vec{x}) \nonumber \\
&=& \zeta_x^{-1}\frac{\sin( \epsilon\pi x)}{\pi x}.
\label{mx}
\end{eqnarray}
Note that, hereafter in this paper, we use the symbol, $m(\vec{x})$, to express the redefined one, $m_{\mathrm{new}}(\vec{x})$ for simplicity.
 At $\vec{x}=\vec{0}$, the value of the right-hand side of Equation (\ref{mx}) is $\zeta_x^{-1}\epsilon$, which is larger than $1$. This leads to the negative value ($\pi$-radian phase shift) of the mask function, $C(1-m(\vec{x}))$. This is fully different from the BLMC that includes only amplitude modulation.
In Figure \ref{concept} (c), we have that the pupil amplitude on the second pupil is exactly nulled using the  new mask. 
Therefore, the Lyot stop is designed such that the aperture of the Lyot stop is exactly the same as the original pupil (Figure \ref{concept}, (d)). 
Also, since $\supp\left[\tilde{m}(\alpha)\right]$ is two times wider than the original pupil aperture, the central peak of $m(x)$ is two times narrower than that of $\tilde{P}(x)$ (Figure \ref{concept}, (g)); therefore, we observe the planets that are closer to the host star.
Furthermore, considering the focal plane, we can express the masked on-axis focal amplitude (see Figure \ref{concept}, (h)) by
\begin{equation}
C(1-m(\vec{x}))\tilde{P}(\vec{x})=C\left(1-\zeta_x^{-1}\frac{\sin ( \epsilon\pi x)}{\pi x}\right)\tilde{P}(\vec{x}).
\label{wavelet}
\end{equation}
Considering the different widths of the central-peak of $m(x)$ and $\tilde{P}(x)$, the function of Equation (\ref{wavelet}) can be interpreted as a (upside-down) wavelet whose Fourier conjugate (the second pupil function) has a non-zero value only outside the original pupil. Meanwhile, the off-axis focus amplitudes are not modulated into such wavelets by the new mask because the positions of those peaks do not match. Therefore, the mask can be said to selectively change the on-axis amplitude to a wavelet. 
\subsection{\label{CON_4th}Fourth-order Null with a Hexagonally Segmented Telescope}
\begin{figure}[H]
\begin{center}
\includegraphics[width=170mm, bb=0 0 1900 800]{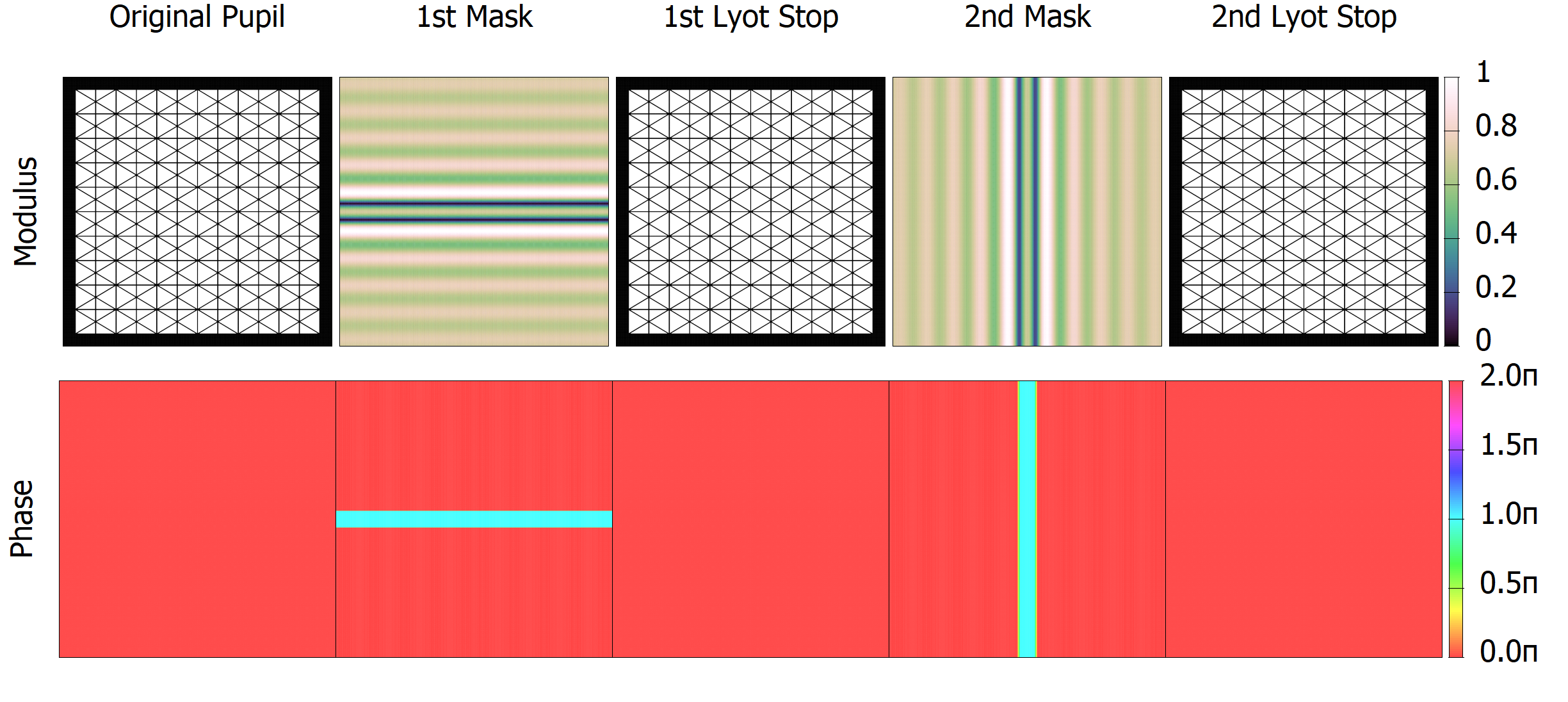}
\caption{\label{colormap} The modulus and phase of the pupil and focus masks used in a fourth-order coronagraphy system. The original pupil in the figure is obtained by the procedure shown in Figure (\ref{remapping}). Two successive new masks are assembled orthogonally to each other with an additional intermediate Lyot stop. The first focus mask is $y$-dependent mask, and the second is $x$-dependent mask.}
\end{center}
\end{figure}

\begin{figure}[H]
\begin{center}
\includegraphics[width=170mm, bb=0 0 2400 1200]{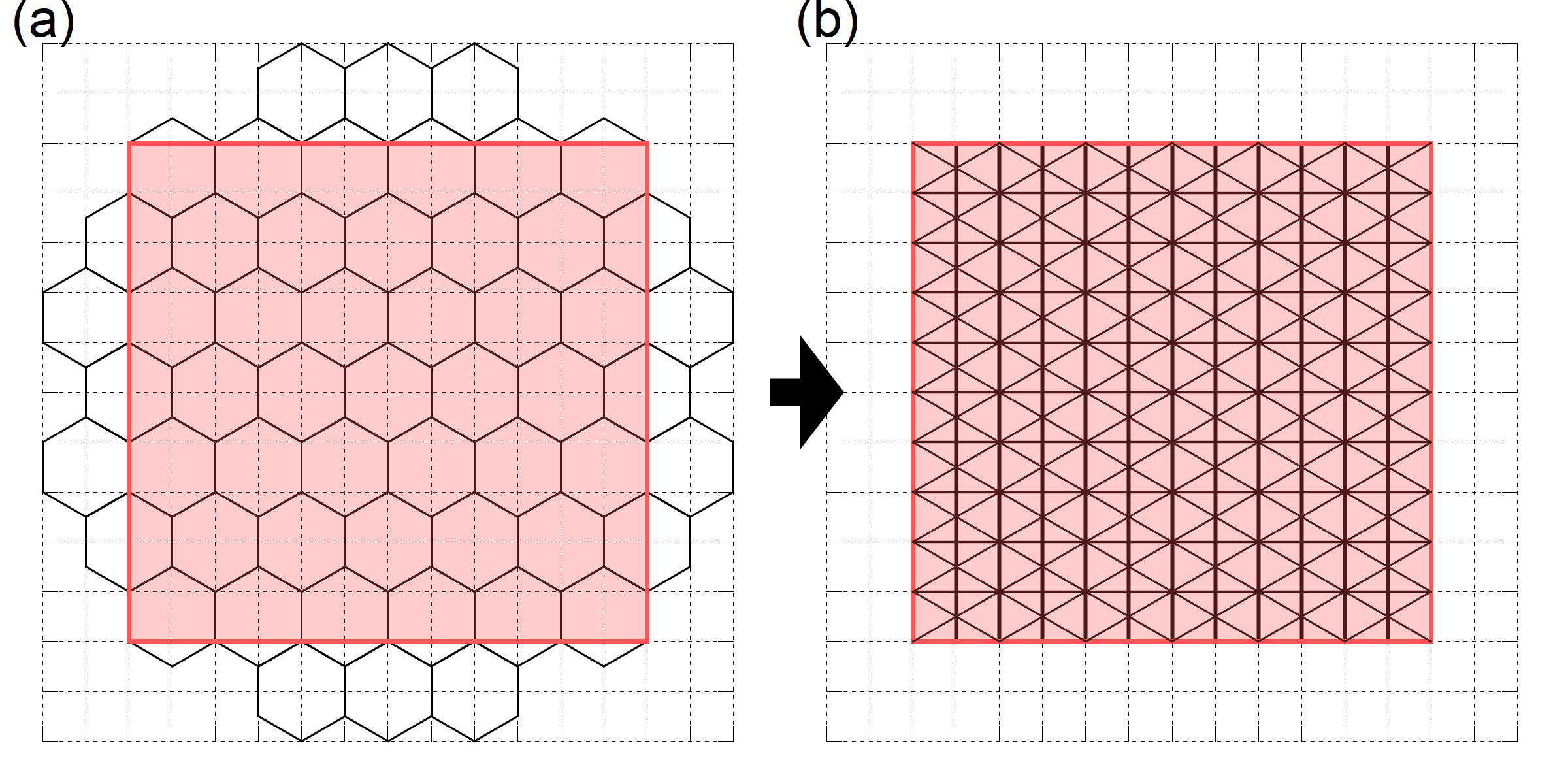}
\caption{\label{remapping}Procedure for making a rectangular pupil optimized for the  new coronagraph . (a)Pupil function of the LUVOIR-b architecture. (b)Optimized pupil function.}
\end{center} 
\end{figure}

It is known that, if we observe a nearby planetary system using a space telescope with a diameter of 5--10m in optical and near-infrared wavelengths, then the second-order null is not sufficient since the angular radius of the target star is typically on the order of $10^{-2} \lambda/D$.
Thus, we introduce a fourth-order null with the  new mask.
To obtain the fourth-order null, we consider the following two points.
(1)Two successive new masks are assembled orthogonally to each other with an additional intermediate Lyot stop. In this paper, we use $y$-dependent mask first, and then $x$-dependent mask (Figure \ref{colormap}). 
(2)The pupil function before employing the focal plane mask must be rectangular aperture with optimized gaps as shown in Figure \ref{remapping}. Also, we required that all the Lyot stops should have the same apertures as the pupil function (Figure \ref{colormap}). 
The optimization of the gaps was carried out such that the pupil satisfies the requirement. The requirement is that both $\zeta_x=\int_{-\infty}^{\infty}\!\! d\alpha P(\vec{\alpha})/P(\vec{\alpha})$ and $\zeta_y=\int_{-\infty}^{\infty}\!\! d\beta P(\vec{\alpha})/P(\vec{\alpha})$ are constant at the original pupil aperture except for the region very close to the gap intersection.
For adequate implementations, we eliminate the defects in areas very close to the gap intersection using a Lyot stop with a shield slightly wider than the gap of the original pupil. 
In this paper, we neglect little changes of the pupil stops, since these changes have little or no impact on the performance of the new mask. 
 The derivation of expressions for performance of the fourth-order system is complied in Appendix \ref{DE}.

The stellar leakage of the fourth-order new mask (normalized by the original intensity of the stellar light) is expressed as below: 
\begin{eqnarray}
L_{4th}(\vec{x},\theta_* ,\gamma ) &=& C^4\left\lbrace \left(\frac{\partial^2}{\partial x\partial y}\tilde{P}(\vec{x})\right)^2 \left( \frac{1}{24}\theta _* ^4+\frac{1}{4}\theta _* ^2\gamma^2+\frac{1}{4}\gamma^4 \right)\right. \nonumber \\
&+&  \left(\frac{\partial^2}{\partial x\partial y}\tilde{P}(\vec{x})\right)A(\vec{x}) \left( \frac{3}{8}\theta _* ^2 \gamma^2 \right) \nonumber \\
&+& \left.A(\vec{x})^2  \left( \frac{1}{32}\theta _* ^4+\frac{3}{16}\theta _* ^2 \gamma^2 +\frac{3}{16}\gamma^4 \right)  \right\rbrace
\label{L4th}
\end{eqnarray}
where $\theta_*$ means stellar angular radius, $\vec{\gamma}$ is normalized angles of telescope pointing jitter, and 
\begin{equation}
A(\vec{x})=4\pi^2\zeta_y^{-1} \int\!\!\!\!\!\!\!\int_{-\infty}^{\infty}\!\!\!\!\!\! d\vec{\alpha}e^{-2\pi i \vec{x}\cdot \vec{\alpha}}P(\vec{\alpha}) \left(1-\zeta_x^{-1}\int_{-\infty}^{\infty}\!\!\!\!\!\! d\alpha \right) P(\vec{\alpha}) \int_{-\infty}^{\infty}\!\!\!\!\!\! d\beta \beta^2 P(\vec{\alpha}).
\end{equation}
The azimuth angle of the tilt direction is set with an uniform distribution of $ [0,2 \pi] $, and Equation (\ref {L4th}) represents the average leakage with respect to the azimuth.
Due to the presence of the intermediate Lyot stop, the leak is not symmetric with respect to the exchange of $x$ and $y$.

\begin{figure}[H]
\begin{center}
\includegraphics[width=140mm, bb=0 0 1200 1440]{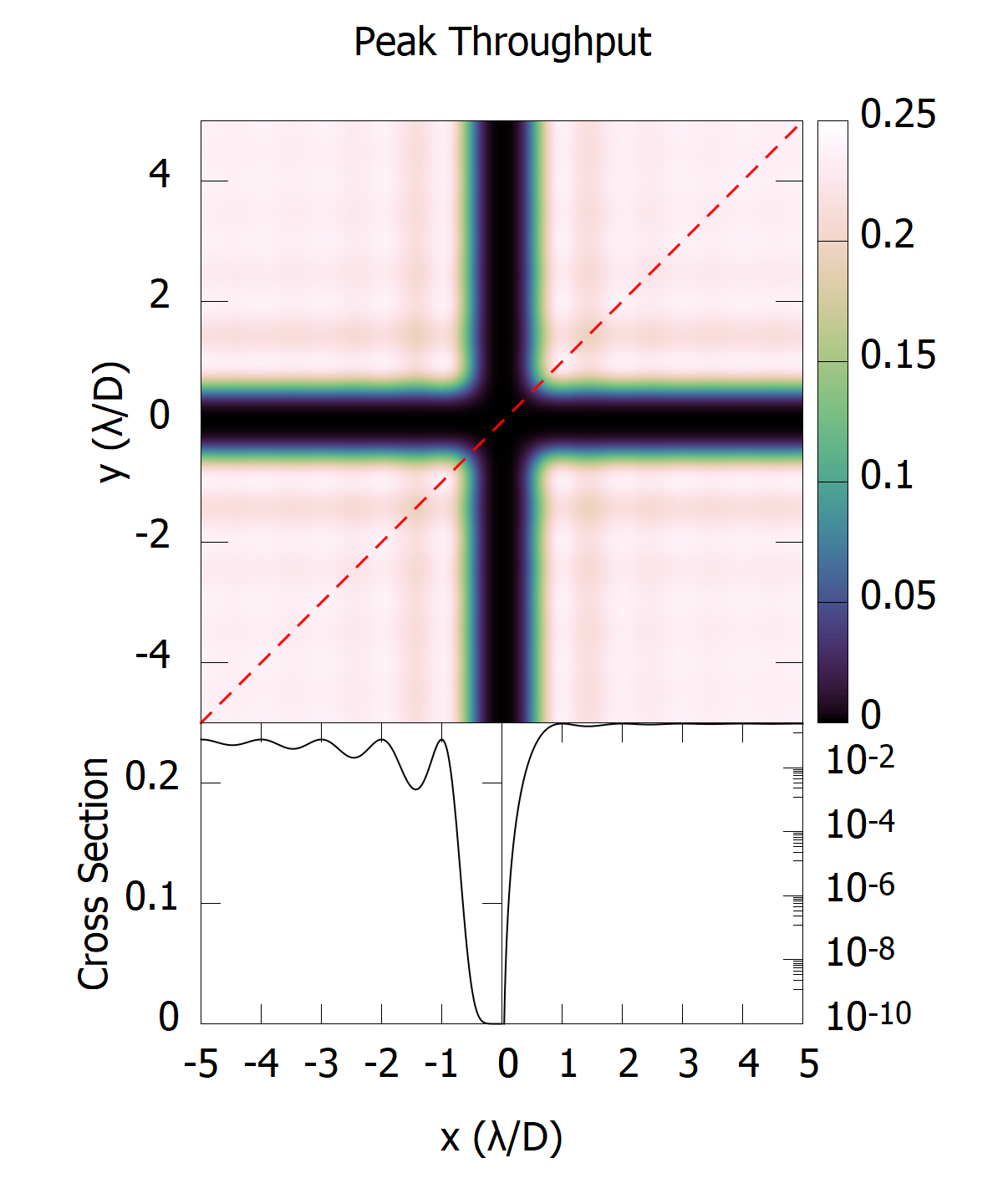}
\caption{\label{TPfig2d}Peak throughput of the fourth-order new coronagraph at off-axis angle ($\epsilon=2$). The lower panel indicates the cross section along the red dashed line ($x=y$) in the upper panel; the lower left and right panels are shown in linear and logarithm. The effect of the constant factor $C^4 (\approx0.25)$ is included. The effect of the pupil function (shown in Figure \ref {remapping}) on the throughput is not considered.}
\end{center}
\end{figure}

The planetary peak throughput is different from the profile of the mask function because a new mask introduces $\pi$-radian-phase modulation.
Figure \ref{TPfig2d} shows the planetary peak throughput of the fourth-order new mask. This peak throughput is evaluated using the equation below:
 \begin{eqnarray}
\tau_{4th}(\vec{\theta}) &=&C^4\left( \left( 1- \left( \frac{\sin( \epsilon\pi x)}{\pi x} \right)^2\right) \left( 1-\left( \frac{\sin( \epsilon\pi y)}{\pi y} \right)^2\right)\right)^2.
\label{TP}
\end{eqnarray}
And since the impact small, we neglect the effect of the pupil gaps on the throughput.
The effect of the obscuration is neglected in Equation (\ref{TP}), see Figure \ref {remapping}, and Figure (\ref{TPfig2d}) shows the result.
Although such light loss independent on $\vec{\theta}$ degrades the ratio of signal to the photon noise, it does not affect the ratio of planetary signals to stellar leaks at all. 
If we neglect such constant factors, then the profile of the peak throughput is close to the theoretical limit proposed by \citet{2006ApJS..167...81G}.

\section{\label{SIM}Simulation}
In this section, the performance of a fourth-order coronagraphic system with an unobscured off-axis segmented telescope is studied and  discussed in terms of detection of the potentially habitable planets.
We carried out the calculation using the analytical expressions of Equations (\ref{L4th}) and (\ref{TP}). 
We consider the monochromatic light of $\mathrm{0.75\mu m}$ and a primary mirror same as that of the LUVOIR-b architecture. 

\subsection{Assumptions}
In the simulation, the normalized angular separation between the planet and its host star and the normalized stellar angular radius with $\lambda/D$,  $\theta_{\mathrm{sep}}(T_{\mathrm{eff}})$ and $\theta_{*}(T_{\mathrm{eff}})$, are assumed as follows:
\begin{equation}
\theta_{\mathrm{sep}}(T_{\mathrm{eff}})=\frac{r_{p}(T_{\mathrm{eff}})}{d}/\left(\lambda/ D\right)
\label{theta_sep}
\end{equation}
\begin{equation}
\theta_{\mathrm{*}}(T_{\mathrm{eff}})=\frac{R_{*}(T_{\mathrm{eff}})}{d}/\left(\lambda/ D\right),
\end{equation}
where $r_{p}(T_{\mathrm{eff}})$ is the geometric mean of the inner and outer edges of the habitable zone proposed by \citet{2013ApJ...765..131K} and $d$ is the distance from a target to the telescope.
Considering the LUVOIR-b architecture as the telescope used for this simulation, $\lambda$ and $D$ are set to $0.75\mathrm{\mu m}$ and $6.4\mathrm{m}$, respectively. 
In Figure \ref{remapping}, we observe that the pupil effective diameter was reduced to $6.4\mathrm{m}$ from the original effective diameter of LUVOIR-b, $8.5\mathrm{m}$. 
The $R_{*}(T_{\mathrm{eff}})$ is the host star radius as a function of the effective temperature. 
We used the empirical relation of \citet{2012ApJ...757..112B} for this function.
Then, we determine the contrast between the stellar and planetary light \citep{2010exop.book..111T} as follow: 
\begin{equation}
\left(\mathrm{Contrast}\right)=\frac{0.367}{\pi}\left(\frac{R_{\oplus}}{r_p}\right)^2,
\end{equation}
where a face-on planet with a radius of $R_{\oplus}$ was considered. 
\subsection{Results}
\begin{figure}[H]
\begin{center}
\includegraphics[width=170mm]{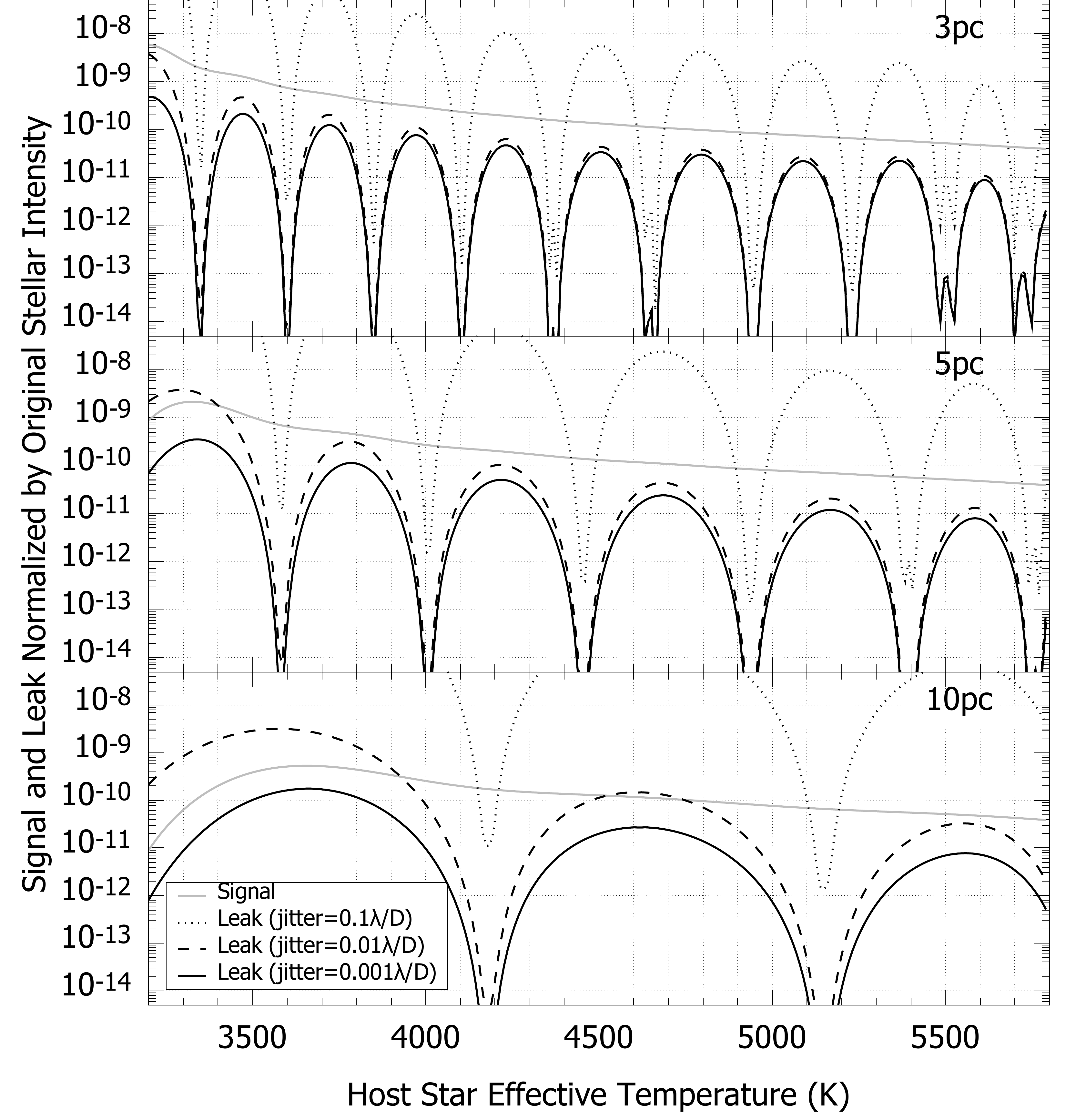}
\caption{\label{simulation}Normalized signals of terrestrial planet in habitable zone by the original stellar intensity (i.e., a product of the contrast between the planet and its host star and the peak throughput) and the stellar leak as a function of various stellar effective temperatures.
The distances from targets to the telescope are set to 3 (top), 5 (center), and 10pc (bottom). The gray solid line shows the planetary signal.  The black dotted, dashed, and solid lines indicate the stellar leak for the telescope pointing jitter of 0.1, 0.01, and 0.001 $\lambda/D$, respectively.}
\end{center}
\end{figure}

Figure \ref{simulation} shows the normalized planetary signal by the original stellar intensity (i.e., the contrast multiplied by the peak throughput), $S(T_{\mathrm{eff}})$, and the stellar leakage, $L(T_{\mathrm{eff}})$.
The results in Figure \ref{simulation} were obtained using Equations (\ref{STeff}) and (\ref{LTeff}).
\begin{equation}
S(T_{\mathrm{eff}})=\left(\mathrm{Contrast}\right) \times \tau_{4th}\left(\frac{\theta_{\mathrm{sep}}(T_{\mathrm{eff}})}{\sqrt{2}}, \frac{\theta_{\mathrm{sep}}(T_{\mathrm{eff}})}{\sqrt{2}}\right)
\label{STeff}
\end{equation}
\begin{equation}
L(T_{\mathrm{eff}})=L_{4th}\left(\frac{\theta_{\mathrm{sep}}(T_{\mathrm{eff}})}{\sqrt{2}}, \frac{\theta_{\mathrm{sep}}(T_{\mathrm{eff}})}{\sqrt{2}},\theta_{*}(T_{\mathrm{eff}}),\gamma \right),
\label{LTeff}
\end{equation}
where $\tau_{4th}(\vec{\theta})$ and $L_{4th}(\vec{\theta})$ are respectively as defined in Equations (\ref{TP}) and (\ref{L4th}). 
The result shows the principle limitation of this coronagraphic performance for monochromatic light of $\mathrm{0.75\mu m}$; 
the higher-order aberration than the telescope pointing jitter was not considered.
The direction of $x=y$ on the image plane was chosen so that the throughput of the planetary signal becomes the largest (see Figure \ref{TPfig2d}).
The multiple peak structures in the plots of the stellar leak is caused by $\left(\frac{\partial^2}{\partial x\partial y}\tilde{P}(\vec{x})\right)^2$ in Equation (\ref{L4th}).
For polychromatic light, the multiple peaks would vanished because the variables of $\left(\frac{\partial^2}{\partial x\partial y}\tilde{P}(\vec{x})\right)^2$ is normalized by $\lambda/D$.
The stellar leakage for the pointing jitter of $0.001 \lambda/D$ is not different from the case of $0.01 \lambda/D$ compared to the difference between the cases of  $0.01 \lambda/D$ and $0.1 \lambda/D$.
This is because the finite stellar angular radius is typically around $0.01 \lambda/D$.
Under the condition that the telescope pointing jitter is $0.001 \lambda/D \approx24\mathrm{\mu as}$, the stellar leakage can be suppressed down to the planetary signal for all types of host stars (3200K--5800K) and distances ($3, 5,$ and $10pc$).
Even for the pointing jitter of $0.01 \lambda/D$, the stellar leakage is approximately less than or equal to the signals of the planets orbiting M-type stars at 3pc, K-type stars at 5pc, and G-type stars at 10pc.
Therefore, this fourth-order coronagraphic system is very effective for detecting the nearby terrestrial planets orbiting G, K, and potentially M-type stars.

\section{\label{DIS}Discussion}
\subsection{Broadband Imaging}
This coronagraphic system should be optimized for a wide observing wavelength range.
However, there is a problem with enlarging its wavelength range because the physical (unnormalized) scale of the point spread function is proportional to the wavelength of the light.
In other words, the profile of the focal plane mask should be scaled by the size of the point spread function.
This means that the performance of the  new coronagraph  system is largely degraded for the broadband imaging. 
To estimate the impact of this factor, we make the following assumption.
Although the mask is designed for the wavelength of $\lambda$, the actual wavelength is $\lambda+\Delta\lambda$, where $\frac{\Delta\lambda}{\lambda}$ is sufficiently smaller than 1. 
 In this case, the spatial scale of the mask function (Figure \ref{concept}, (g)) is magnified by $\frac{\lambda}{\lambda+\Delta\lambda}$ from the appropriate scale.
Therefore, non-zero values of the Fourier conjugate of the mask function (Figure \ref{concept}, (b)), excluding the central delta function peak, are scaled by $\frac{\lambda}{\lambda+\Delta\lambda}\approx 1-\left(\frac{\Delta\lambda}{\lambda}\right)$ from the appropriate value.
Hence, the remaining modulus of the amplitude on the second pupil is constant over the original pupil and is approximately estimated as  $\left(\frac{\Delta\lambda}{\lambda}\right)$.
For the fourth-order version, the remaining modulus of the amplitude on the last pupil becomes the square of the second-order version,  $\left(\left(\frac{\Delta\lambda}{\lambda}\right)\right)^2=\left(\frac{\Delta\lambda}{\lambda}\right)^2$.
Thus, the leaks caused by the deviation of the wavelength is obtained as $\left(\frac{\Delta\lambda}{\lambda}\right)^4$ times the point spread function before the coronagraphic system.
This means that $\frac{\Delta\lambda}{\lambda}$ must be smaller than  0.32\% for the contrast of $10^{-10}$.
Hence, we desire a method for implementing an achromatic  new mask. 
\subsection{Prospects for Manufacturing}
We also discuss the prospects for manufacturing the pupil stops and focal plane masks of the  new coronagraph .

We assume that the manufacturing errors of the first pupil stop size and the first Lyot stop size are  $\sigma_1$ and $\sigma_2$  in their standard deviation; these errors are normalized by the pupil full width.
In this case, the on-axis source leakage caused by the manufacturing errors is proportional to $\sigma_1^2 \sigma_2^2 $, where the errors are assumed to be the Gaussian distributions independent of each other.
Therefore, we estimate the manufacturing tolerance for the pupil stop size to be $10^{-10/4}\approx0.3\%$ of the pupil full width when aiming at the contrast of $10^{-10}$.  

We also assume that the manufacturing errors of the first and second focal plane masks in terms of amplitude transmittance are $e_1(\vec{x})$ and $e_2(\vec{x})$, respectively.
We cannot simply estimate the resultant leakage of the fourth-order coronagrahic system at each point on the focal plane, $\left|e(\vec{x})\right|^2$, since it depends on the functional forms of  $e_1(\vec{x})$ and $e_2(\vec{x})$.
However, the upper limit of $\iint ^{\infty}_{-\infty} d\vec{x} \left|e(\vec{x})\right|^2$ is obtained as follows:
\begin{equation}
 \iint ^{\infty}_{-\infty} d\vec{x} \left|e(\vec{x})\right|^2 \leq \prod_{i} \iint ^{\infty}_{-\infty} d\vec{x} \left|e_i(\vec{x})\right|^2.
\end{equation}

In the actual manufacturing, the focal plane masks of the  new coronagraph  is made as notch-filter masks \citep{2003ApJ...594..617K,2004ApJ...608.1095D}.
This is because the notch filter mask is one of the binary masks whose Fourier conjugate is equal to the Fourier conjugate of the band-limited mask in the region two times wider than the Lyot stop.
The difference between these Fourier conjugates has no impact on the convolution in Equation (\ref{convolution}).

\section{Conclusion}
In this paper, we proposed a coronagraphic system with fourth-order null suitable for off-axis unobscured  segmented telescopes.
 The coronagraphic system is comprised of a new type of focal plane mask and a Lyot stop that is almost the same as the original pupil.
The new system is sufficiently insensitive to the telescope pointing errors and finite host star angular diameter thanks to the fourth-order null of the host star.
 Also, this coronagraphic system can suppress the stellar halo over a wide range of angular separations.
The inner working angle is close to $1\lambda/D$ and there are no outer limits. 
The Fourier conjugate of the  new mask function is the sum of the delta function and $-1$ time the rectangular function that has twice the width of the original pupil. 
Therefore, despite the application of the coronagraphic system to an off-axis unobscured segmented telescope, we can null the amplitude at the re-imaged pupil. 
The  new coronagraph  achieves the second-order null.
The fourth-order version of the  new coronagraph is constructed by placing two new masks in succession to be orthogonal to each other.
 We studied and investigated the performance of this fourth-order band-limited mask coronagraph for observations in monochromatic $\lambda=0.75\mathrm{\mu m}$  light with an 8.5-meter segmented telescope.
According to the result, potentially habitable planets around nearby G and K-type main sequence stars can be directly detected for the pointing jitter less than $0.01 \lambda/D$ or less of the pointing jitter. 

We estimated the limited bandwidth due to the contrast of $10^{-10}$;  $\frac{\Delta\lambda}{\lambda}$ must be smaller than $0.32\%$ for the contrast of $10^{-10}$.
The manufacturing tolerance for the pupil stop size is $10^{-10/4}\approx0.3\%$ of the pupil full width for the contrast of $10^{-10}$.
The focal plane masks of the  new coronagraph  would be possibly made as notch-filter masks.
However, further investigation of this mask is highly encouraged to make an absolute statement.
The new coronagraph mask concept is effective for the direct imaging of potentially habitable planets with the future space off-axis segmented telescopes.
\bibliographystyle{apj}
\bibliography{ms}
\appendix
\renewcommand{\theequation}{A.\arabic{equation}}
\setcounter{equation}{0}
\section{\label{DE}Derivation of Expressions for Performance of the Fourth-order System}
 This appendix derives the characteristics of the fourth-order coronagraph system proposed in Section \ref{CON_4th}.
In the following, the aperture function of the pupil just before the focal-plane mask, the first and second Lyot stops are denoted $P_{1}(\vec{\alpha})$, $P_{2}(\vec{\alpha})$ and $P_{3}(\vec{\alpha})$ respectively.
The values of these aperture functions can be $1$ or $0$. 
The followings are assumed:
\begin{equation}
P_{2}(\vec{\alpha})=P_{2}(\vec{\alpha})P_{1}(\vec{\alpha}),
\end{equation}
\begin{equation}
P_{3}(\vec{\alpha})=P_{3}(\vec{\alpha})P_{2}(\vec{\alpha})P_{1}(\vec{\alpha}).
\end{equation}
Additionally, the first and second masks, $M_{y}(\vec{x})$ and $M_{x}(\vec{x})$, are defined as follows:
\begin{equation}
M_{y}(\vec{x})=C\left(1-\zeta_y^{-1}\frac{\sin ( \epsilon\pi y)}{\pi y}\right),
\label{mask_y}
\end{equation}
\begin{equation}
M_{x}(\vec{x})=C\left(1-\zeta_x^{-1}\frac{\sin ( \epsilon\pi x)}{\pi x}\right),
\label{mask_x}
\end{equation}
where $2\leq\epsilon$. 
In the followings, the variables of these functions such as $\vec{\alpha}$ or $\vec{x}$ are sometimes omitted  for notational simplicity.

We first derive the point spread function for the fourth-order coronagraph system for preparation.
The point spread function that depends on the position angle of the point source, $\vec{\theta}$, can be derived as follows:
\begin{eqnarray}
\mathrm{PSF}(\vec{x},\vec{\theta})&=&C^4\left| \int\!\!\!\!\!\!\!\int_{-\infty}^{\infty}\!\!\!\!\!\! d\vec{\alpha}e^{-2\pi i \vec{x}\cdot \vec{\alpha}} P_{3}\left(\tilde{M_{x}}\ast \left( P_{2}\left( \tilde{M_{y}} \ast\left( P_{1}e^{2\pi i \vec{\theta}\cdot \vec{\alpha}}\right)\right)\right)\right) \right|^2 \nonumber \\
&=&C^4\left| \int\!\!\!\!\!\!\!\int_{-\infty}^{\infty}\!\!\!\!\!\! d\vec{\alpha}e^{-2\pi i \vec{x}\cdot \vec{\alpha}} P_{3}\left(1-\zeta_{x}^{-1}\int_{-\infty}^{\infty}\!\!\!\!\!\! d\alpha \right) P_{2}\left(1-\zeta_{y}^{-1}\int_{-\infty}^{\infty}\!\!\!\!\!\! d\beta \right) P_{1}e^{2\pi i \vec{\theta}\cdot \vec{\alpha}}\right|^2 .
\end{eqnarray}
When $P_1=P_2=P_3=\rect\left[ \alpha \right]\rect\left[ \beta \right]$, $\mathrm{PSF}(\vec{x},\vec{\theta})$ can be simply expressed by
\begin{eqnarray}
\mathrm{PSF}(\vec{x},\vec{\theta})&=&\mathrm{PSF}_{1-D}(x,\theta_x)\mathrm{PSF}_{1-D}(y,\theta_y),
\label{ppssff}
\end{eqnarray}
where 
\begin{eqnarray}
\mathrm{PSF}_{1-D}(x,\theta_x)&=&C^2 \left| \left(\frac{\sin ( \epsilon\pi (x-\theta_x ))}{\pi (x-\theta_x )}-\frac{\sin ( \epsilon\pi x)}{\pi x}\frac{\sin ( \epsilon\pi \theta_x)}{\pi \theta_x}\right) \right| ^2.
\label{ppssff1d}
\end{eqnarray}

\subsection{Peak Throughput}
To derive the peak throughput of the system, substitute $\vec{x}=\vec{\theta}$ into Equations (\ref{ppssff}) and (\ref{ppssff1d}).
\begin{eqnarray}
\tau_{4th}(\vec{\theta}) &=& \mathrm{PSF}(\vec{\theta},\vec{\theta}) \nonumber \\
&=& C^4\left| \left(1-\left(\frac{\sin ( \epsilon\pi \theta_x )}{\pi \theta_x }\right)^2\right) \left(1-\left(\frac{\sin ( \epsilon\pi \theta_y )}{\pi \theta_y }\right)^2\right)\right|^2.
\end{eqnarray}

\subsection{Leak of Stellar Light}
 
In the followings, we assume 
\begin{equation}
P_3=P_2=P_1=P
\end{equation}
for simplicity.
$P$ is a real function thus $\tilde{P}$ is an even function. 
Using this fact, Taylor-expansion of  $\mathrm{PSF}(\vec{x},\vec{\theta})$ about $\vec{\theta}$ around $\vec{\theta}=\vec{0}$ can be calculated as follows:
\begin{eqnarray}
\mathrm{PSF}(\vec{x},\vec{\theta})&=&C^4\left( A_{xy}(\vec{x})^2 \theta_x^2 \theta_y^2+A_{xy}(\vec{x})A_{yy}(\vec{x}) \theta_x \theta_y^3 + \frac{A_{yy}(\vec{x})^2}{4} \theta_y^4 \right),
\label{tay}
\end{eqnarray}
where terms of higher order than 4 are ignored.
In Equation (\ref{tay}), $A_{xy}$ and $A_{yy}$ are defined as follows:
\begin{equation}
A_{xy}(\vec{x})=\frac{\partial^2}{\partial x\partial y} \tilde{P}(\vec{x}),
\end{equation}
\begin{equation}
A_{yy}(\vec{x})=4\pi^2\zeta_y^{-1} \int\!\!\!\!\!\!\!\int_{-\infty}^{\infty}\!\!\!\!\!\! d\vec{\alpha}e^{-2\pi i \vec{x}\cdot \vec{\alpha}}P(\vec{\alpha}) \left(1-\zeta_x^{-1}\int_{-\infty}^{\infty}\!\!\!\!\!\! d\alpha \right) P(\vec{\alpha}) \int_{-\infty}^{\infty}\!\!\!\!\!\! d\beta \beta^2 P(\vec{\alpha}).
\end{equation}
We use Equation (\ref{tay}) and the following expression to calculate the stellar leakage of the fourth-order  new mask (normalized by the original intensity of the stellar light): 
\begin{eqnarray}
L_{4th}(\vec{x},\theta_* ,\gamma ) &=&\frac{1}{2\pi}\int_{0}^{2\pi}\!\!\! d\Phi \frac{1}{\pi \theta_* ^2} \int_{0}^{\theta_*}\!\!\!\!\theta d\theta \int_{0}^{2\pi}\!\!\! d\phi \mathrm{PSF}(\vec{x},\vec{\theta}-\vec{\gamma}),
\label{ll4}
\end{eqnarray}
where $\vec{\gamma}$ means pointing jitter, 
\begin{equation}
\vec{\gamma}=\left( \gamma \cos(\Phi), \gamma \sin(\Phi) \right),
\end{equation}
and the operator, $\frac{1}{2\pi}\int_{0}^{2\pi}\!\!\! d\Phi $, expresses the average with respect to the azimuth, $\Phi$.
To calculate the right-hand side of Equation (\ref{ll4}), we define the following definite integrals:
\begin{eqnarray}
L_{nm}&=& \int_{0}^{2\pi}\!\!\! d\phi \left(\cos(\phi)\right)^n \left(\sin(\phi)\right)^m,
\label{Lnm}
\end{eqnarray}
where $n$ and $m$ are integers.
When both $n$ and $m$ are even numbers,
\begin{eqnarray}
L_{nm}&=& L_{mn} \nonumber \\
&=& 2^{1-n-m}\pi \sum_{k=0}^{n}\binom{n}{k}\binom{m}{\frac{n+m}{2}-k} \left( -1\right) ^k.
\end{eqnarray}
When either $n$ or $m$ are odd numbers,
\begin{eqnarray}
L_{nm}&=& 0.
\end{eqnarray}
The values for even $(n,m)$ up to 4 are as follows:
 \begin{eqnarray}
\left( L_{00},L_{20},L_{22},L_{40}\right)&=& \left( 2\pi,\pi,\frac{\pi}{4},\frac{3\pi}{4}\right).
\end{eqnarray}
Using these definite integrals,  the right-hand side of Equation (\ref{ll4}) can be calculated as follows:
\begin{eqnarray}
L_{4th}(\vec{x},\theta_* ,\gamma ) &=& \frac{C^4}{\pi\theta_{*}^2} \frac{1}{2\pi}\int_{0}^{2\pi}\!\!\! d\Phi \int_{0}^{\theta_*}\!\!\!\!\theta d\theta\left\lbrace A_{xy}(\vec{x})^2\left( L_{22}\theta ^4+L_{20}\theta ^2\gamma ^2+L_{00}\gamma_x ^2\gamma_y ^2 \right) \right. \nonumber \\ 
&+&  A_{xy}(\vec{x})A_{yy}(\vec{x})\left( 3L_{02}\theta ^2\gamma _x ^2+L_{00}\gamma_x \gamma_y ^3 \right) \nonumber \\ 
&+&  \left.\frac{1}{4}A_{yy}(\vec{x})^2\left( L_{04}\theta ^4+6L_{02}\theta ^2\gamma _y ^2+L_{00} \gamma_y ^4 \right) \right\rbrace \nonumber \\
&=& C^4\frac{1}{2\pi}\int_{0}^{2\pi}\!\!\! d\Phi \left\lbrace A_{xy}(\vec{x})^2 \left( \frac{1}{24}\theta_{*}^4 + \frac{1}{4} \gamma^2 \theta_{*}^2 + \gamma_x ^2\gamma_y ^2 \right) \right. \nonumber \\ 
&+&  A_{xy}(\vec{x})A_{yy}(\vec{x})\left( \frac{3}{4}\theta_{*} ^2\gamma _x ^2+\gamma_x \gamma_y ^3 \right) \nonumber \\ 
&+&  \left.A_{yy}(\vec{x})^2\left( \frac{1}{32}\theta_{*} ^4+ \frac{3}{8}\theta_{*} ^2\gamma _y ^2+\frac{1}{4}\gamma_y ^4 \right) \right\rbrace \nonumber \\
&=& C^4  \left\lbrace A_{xy}(\vec{x})^2 \left( \frac{1}{24}\theta_{*}^4  + \frac{1}{4} \gamma^2 \theta_{*}^2 +\gamma ^4 \frac{L_{22}}{2\pi}\right) \right. \nonumber \\ 
&+&  A_{xy}(\vec{x})A_{yy}(\vec{x})\left( \frac{3}{4}\theta_{*} ^2\gamma ^2 \frac{L_{02}}{2\pi} \right) \nonumber \\ 
&+&  \left.A_{yy}(\vec{x})^2\left( \frac{1}{32}\theta_{*} ^4+ \frac{3}{8}\theta_{*} ^2\gamma ^2\frac{L_{02}}{2\pi}+\frac{1}{4}\gamma ^4 \frac{L_{04}}{2\pi} \right) \right\rbrace \nonumber \\
&=& C^4  \left\lbrace A_{xy}(\vec{x})^2 \left( \frac{1}{24}\theta_{*}^4  + \frac{1}{4} \gamma^2 \theta_{*}^2 +\frac{1}{4}\gamma ^4 \right) \right. \nonumber \\ 
&+&  A_{xy}(\vec{x})A_{yy}(\vec{x})\left( \frac{3}{8}\theta_{*} ^2\gamma ^2  \right) \nonumber \\ 
&+&  \left.A_{yy}(\vec{x})^2\left( \frac{1}{32}\theta_{*} ^4+ \frac{3}{16}\theta_{*} ^2\gamma ^2+\frac{3}{16}\gamma ^4  \right) \right\rbrace.
\end{eqnarray}

\newpage

\begin{table}[H]
\begin{center}
\begin{tabular}{c c}
\hline \hline
Symbol & Meaning \\
\hline
$\lambda$ & wavelength of light \\	
\hline
$D$ & effective diameter of telescope \\
 \hline
$D_{\alpha}, D_{\beta}$ & $\alpha$- or $\beta$- directional aperture full width of telescope \\
 \hline
$\vec{x}=(x,y)$ & focal-plane angular coordinate normalized by $\lambda/D$ \\
\hline
$\vec{\alpha}=(\alpha,\beta)$ & pupil-plane coordinate normalized by $D$ \\
\hline
$\gamma$ & absolute value of the telescope's pointing deviation \\
\hline
$\delta(x)$ & Dirac's delta function \\
\hline
$\rect(x)$ &  $\rect(x) = \left( 1\ \ (|x|\leq0.5),\ 0.5\ \ (|x|=0.5),\ 0\ \ (|x|\geq1) \right) $ \\
\hline
$f*g$ & convolution of functions, $f$ and $g$ \\
\hline
$\tilde{f}$ & Fourier conjugate of a function, $f$ \\
\hline
$\mathrm{Supp}(f)$ & $\left\lbrace \vec{x}|f(\vec{x})\neq 0 \right\rbrace$ \\
\hline
$P(\vec{\alpha})$ & original pupil function \\
\hline
$M(\vec{x})$ & mask function \\
\hline
$C$ & constant factor in mask function \\
\hline
$m(x)$ & $1-M(\vec{x})/C$ \\
\hline
$\epsilon$ & mask parameter \\
\hline
$\theta_{*}$ & stellar angular radius \\
\hline
$L_{4th}(\vec{x},\theta_{*},\vec{\gamma})$ & leakage of light from central star \\
\hline
$\tau_{4th}(\vec{\theta})$ & peak throughput \\
\hline
$\zeta_x, \zeta_y$ & constants that correspond to $\int_{\infty}^{\infty}\!\! d\alpha P(\vec{\alpha})$ and $\int_{\infty}^{\infty}\!\! d\beta P(\vec{\alpha})$ at $\supp\left[P\right]$  \\
\hline
$d$ & distance from target to the telescope \\
\hline
$\theta _{\mathrm{sep}} (T_{\mathrm{eff}})$ & angular separation of habitable planets normalized by $\lambda/D$ \\
\hline
$\theta_ {*} (T_{\mathrm{eff}})$ & angular radii of host stars normalized by $\lambda/D$ \\
\hline
$r_{p}(T_{\mathrm{eff}})$ & \begin{tabular}{c}geometric mean of the inner and outer edges\\ of the habitable zone proposed by \citet{2013ApJ...765..131K}. \end{tabular}\\
\hline
\hline
\end{tabular}
\end{center}
\caption{Notation of symbols}
\label{notation}
\end{table}

\end{document}